\documentclass[
  journal=medium,
  manuscript=article-type,
  year=XXXX,
  volume=XX,
]{cup-journal}

\usepackage[utf8]{inputenc}
\usepackage{amsmath}
\usepackage[nopatch]{microtype}
\usepackage{booktabs}
\usepackage{hyperref}
\usepackage{amsmath, amsfonts, amssymb, amsthm}
\usepackage{comment}
\usepackage{caption}
\usepackage{subcaption}

\DeclareUnicodeCharacter{0301}{\'{e}}

\newcommand{\bfalp}{\boldsymbol{\alpha}}
\newcommand{\bfbet}{\boldsymbol{\beta}}
\newcommand{\bfSig}{\boldsymbol{\Sigma}}
\newcommand{\bfrho}{\boldsymbol{\rho}}
\newcommand{\bfx}{\boldsymbol{x}}
\newcommand{\bfX}{\boldsymbol{X}}
\newcommand{\bfxi}{\boldsymbol{\xi}}
\newcommand{\bfeta}{\boldsymbol{\eta}}
\newcommand{\bfzeta}{\boldsymbol{\zeta}}

\newcommand{\alf}{\alpha}
\newcommand{\bet}{\beta}
\newcommand{\eps}{\epsilon}
\newcommand{\del}{\delta}
\newcommand{\gam}{\gamma}

\newcommand{\Sig}{\Sigma}

\newcommand{\kap}{\kappa}

\newcommand{\simiid}{{\,\, \overset{\text{iid}}\sim} \,\,}

\newcommand{\E}{\mathbb{E}}
\newcommand{\reals}{\mathbb{R}}

\newcommand{\Ber}{\mathsf{Ber}}
\newcommand{\normal}{\mathsf{N}}

\title{Explaining Differences in Voting Patterns Across Voting Domains Using Hierarchical Bayesian Models}

\author{Erin Lipman}
\affiliation{Department of Statistics, University of Washington, Seattle, WA, USA}
\email[Erin Lipman]{erlipman@uw.edu}

\author{Scott Moser}
\affiliation{School of Politics and International Relations, University of Nottingham, Nottingham, UK}

\author{Abel Rodriguez}
\affiliation{Department of Statistics, University of Washington, Seattle, WA, USA}

\keywords{roll call votes, hierarchical model, U.S.\ House of Representatives} 

\begin{document}

\begin{abstract}
Spatial voting models of legislators' preferences are used in political science to test theories about their voting behavior. These models posit that legislators' ideologies as well as the ideologies reflected in votes for and against a bill or measure exist as points in some low dimensional space, and that legislators vote for positions that are close to their own ideologies. Bayesian spatial voting models have been developed to test sharp hypotheses about whether a legislator's revealed ideal point differs for two distinct sets of measures. This project extends such a model to identify covariates that explain whether legislators exhibit such differences in ideal points. We use our method to examine voting behavior on procedural versus final passage votes in the U.S. house of representatives for the 93\textsuperscript{rd} through 113\textsuperscript{th} congresses. The analysis provides evidence that legislators in the minority party as well as legislators with a moderate constituency are more likely to have different ideal points for procedural versus final passage votes.
\end{abstract}

\section{Introduction}


Estimating members' preferences from roll-called votes in parliaments and legislatures is crucial for theory-testing in political science \citep{kreh90-are,ther06-party}. Significant progress has been made in the methodology of ideal point estimation since the seminal work of \textcite{pool85-spatial} and \textcite{pool00-congress}, with early contributions from \textcite{aldr77-method} and others. However, methodological challenges remain when comparing ideal points over time, across domains, or between chambers. Many of these challenges have been addressed in the literature (e.g., \citealp{poole2005spatial}, \citealp{shor2010bridge}, \citealp{jess16-how}), but others remain, especially in settings where the interest is theory testing. In that kind of setting, there are two related but distinct sets of problems that need to be addressed: \emph{scaling} ideal points to allow for comparisons and \emph{explaining} variation in estimated ideal points.  The focus of this paper is on devising methodology that enables both tasks to be carried out simultaneously.


More concretely, we are interested in settings where legislators vote on motions in various domains and where they might exhibit different revealed preferences across domains. These domains could be, for example, issue areas (e.g., defense, housing, government operations, etc., as in \citealp{jone05-politics} or \citealp{moser2021multiple}), 
dimensions (e.g., social vs.\ economic, as in \citealp{pool00-congress,pool11-ideology}), or type of vote (e.g., procedural vs.\ final passage, as in  \citealp{jess14-two}).\footnote{There are   many other ways to categorise roll-called votes, e.g., 
Clausen categories \citep{clausen1973policy}, Peltzman categories, and Specific Issue Codes.  For example, see   \url{https://voteview.com/articles/issue_codes}.} Legislators may have different revealed preferences across these domains. For instance, \textcite{jess14-two} found that party polarization is greater in the procedural domain, meaning that when using roll call data only from votes on procedural motions, party polarization is greater than when legislators vote on amendments or final passage motions. This pattern is similar to the findings of \textcite{kirk17-ideology} when looking at defections in the minority party.  To ensure common scaling of ideal points across domains, previous literature assumed that a small number of legislators (often referred to as ``bridges'') had the same revealed preferences across domains \citep{shor2010bridge,shor11-ideological,trei11-comparing}. This assumption was relaxed by \textcite{Lof16_u} and \textcite{moser2021multiple}, who estimated the identity of the bridges. These models produce a posterior distributions for the identity of each legislator as a bridge (the expected bridge probability) and, in the case of more than two voting domains, a clustering of legislators with similar patterns of changing revealed preferences across domains. 

While the previous models enable us to identify bridges, they do not directly allow us to incorporate covariates that can potentially explain changes in revealed preferences across domains.  Understanding factors that influence voting behavior in different domains and changes in voting patterns across domains has the potential to contribute to studies about party influence on legislative voting \citep{snyd01-estimating}, dynamics of polarization \citep{jess14-two,sulk14-partisan}, and party defection \citep{fino99-breaking}. Additionally, answering this question sheds light on the broader literature on determinants of legislative voting. Such studies typically either treat all votes the same (e.g., \citealp{pool00-congress}) or hand-pick particular domains in which to use roll-call votes (e.g. \citealp{jeon09-constituent,muri22-heeding}). Our approach extends these approaches by allowing for domain-specific voting patterns and simultaneously estimating factors that influence changes in voting behavior across the two domains. 

This paper develops a fully Bayesian hierarchical model that jointly addresses the questions of how to properly scale ideal points across two voting domains in order to estimate the identity of the bridge legislators and of how to identify factors that might explain the fact that preferences vary across domains.  This allows for testing hypotheses such as ``Does covariate \(X\) matter for explaining differences in voting patterns across domains?'' For example, our approach allows for testing claims such as ``constituency characteristics have no effect on the probability that a member will exhibit different revealed preferences across domains'' or ``being in the minority party has no effect on the probability of having the same revealed preference across all domains.''  The use of a joint model has a number of advantages over a two-step procedure, in which a point estimate of the bridge identities is first estimated from the voting data and then a regression model is fitted using these point estimates as the response variables.  Two-step procedures are potentially sensitive to the way the point estimator of the bridge identities is constructed (e.g., the threshold used for the posterior probabilities that an individual legislator is a bridge).  Furthermore, because the ideal points of the legislators and the identity of the bridges are estimated from the voting records, and therefore and subject to uncertainty, two-step procedures can be expected to underestimate the uncertainty associated with the variable selection step. Joint models avoid both of these issues.  

We illustrate our joint model through a study of the factors that explain why legislators vote differently on procedural and final passage votes in the modern U.S.\ House of Representatives.  As we already discussed, the literature views procedural votes separately from substantive ones (e.g., see \citealp{ther06-party}, \citealp{jess14-two}, \citealp{patt10-dilatory}, \citealp{cars14-procedural,good04-lameduck} and \citealp{kirk14-partisanship}).  Our empirical analysis builds on this literature by simultaneously considering a panel of constituency-level, legislator-level, and chamber-level factors over an extended period of 42 years ending in 2014.

The remainder of the paper is organized as follows. Section \ref{sec:methods} presents describes the model and discusses its computational implementation and the identifiability of model parameters. Section \ref{se:example} applies the model to voting in the U.S.\ House of Representatives.  Finally, Section \ref{se:discussion} discusses potential shortcomings of our approach and future research directions.

\section{The Model}\label{sec:methods}

\subsection{Spatial voting in multiple domains}

Let $y_{i,j}\in \{0,1\}$ correspond to the vote of legislator $i=1,\dots,I$ on measure $j=1,\dots,J$, with $y_{i,j} = 0$ representing a negative (``nay'') vote and $y_{i,j}=1$ representing a positive (``yay'') vote. In the spirit of \citet{jackman2001multidimensional} and \citet{clinton2004statistical}, 
we assume that 
\begin{equation}\label{eq:likelihood2}
y_{i,j} \mid \mu_j, \bfalp_j, \bfbet_{i,0}, \bfbet_{i,1}, \gamma_j \sim \Ber \left(y_{i,j} \, \Bigg | \, 
%
\frac{1}{1 + \exp \left\{ -\left(\mu_j+\bfalp_j^T\bfbet_{i,\gam_j} \right)\right\}}
\right),
\end{equation}
where $\mu_j \in \reals$ and $\bfalp_j \in \reals^d$ are unknown parameters that can be interpreted, respectively, as the baseline probability of an affirmative vote and the discrimination associated with measure $j$, $\gamma_j \in \{ 0,1\}$ is a known indicator variable of whether the $j$-th vote corresponds to a procedural ($\gamma_j = 0$) or final passage ($\gamma_j = 1$) vote, $\bfbet_{i,0} , \bfbet_{i,1} \in \reals^d$ are unknown parameters representing the ideal point of legislator $i$ on procedural and final passage votes, respectively, and $d$ is the (maximum) dimension of the latent policy space, which is assumed to be known.  This statistical model can be derived from a spatial voting model \citep{enelow1984spatial,davis1970expository} using the random utility framework of \textcite{mcfadden1973conditional}.  It also mimics the structure of a logistic item response theory (IRT) model \citep{fox2010bayesian}.  However, unlike traditional IRT models, it allows for each legislator to have (in principle) different ideal points on each of the two voting domains.

In this paper we adopt a Bayesian approach to inference.  Therefore, we need to specify priors for the model parameters.  Our approach to prior elicitation is similar to that in \textcite{bafumi13andrew}, who advocate the use of hierarchical priors.  In particular, for the intercepts $\mu_1, \ldots, \mu_J$, we let 
\begin{align*}
\mu_j \mid \rho_\mu, \kap^2_\mu & \simiid \normal(\mu_j \mid \rho_\mu, \kap^2_\mu), & j&=1, \ldots, J,
\end{align*}
where $\normal(\cdot \mid a, b^2)$ denotes the normal distribution with mean $a$ and variance $b^2$. The hyperparameters $\rho_\mu$ and $\kap^2_\mu$ are then given normal and inverse gamma hyperpriors respectively. On the other hand, for the discrimination parameters $\bfalp_1, \ldots, \bfalp_J$ we set 
\begin{align*}
\alf_{j,k} \mid \omega_\alf, \kap^2_\alf  &\simiid \omega_{\alf,k} \del_{0}(\alf_{j,k}) 
+ (1-\omega_{\alf,k})\normal(\alf_{j,k} \mid 0,\kap^2_\alf), & j&=1,\ldots, J , & k=1,\ldots, d,
\end{align*}
where $\delta_0$ is a point mass at $0$. This prior is completed by assigning an inverse gamma hyperprior on $\kappa_{\alpha}^2$ and independent beta hyperpriors on $\omega_{\alf,1}, \ldots, \omega_{\alf,d}$.  The use of zero-inflated Gaussian distributions for the discrimination parameters 
allows us to automatically handle unanimous votes without having to explicitly remove them from the dataset.

\subsection{Testing sharp hypothesis about differences in ideal points}

To complete our model, we need to assign prior distributions to the ideal points of each legislator and voting domain (in our application, procedural or final passage voting).  The simplest alternative, independent priors across both legislators and types, has several shortcomings.  Most importantly, such an approach is (roughly) equivalent to fitting independent IRT models of the kind described in \textcite{jackman2001multidimensional} for each group of votes.  Under such an approach, the two latent spaces are unlinked and are therefore incomparable.  Indeed, while an absolute scale can be created for each one of them independently by, for example, fixing the position of a small number of legislators on each of the scales separately, the only way to ensure comparability is to assume we know how the position of these legislators changes (or not) when moving from voting on procedural to final passage motions.  While assumptions of this kind have been successfully used in the past (for example, by \citealp{shor2010bridge} to create a common scale to map the relative ideological positions of various state legislatures),  they are very strong.  In particular, some of the natural ``bridge'' legislators (such as the party leaders) are those whose behavior might be most interesting to study.

Hence, in this paper we adopt an approach similar that of \textcite{Lof16_u} and elicit a joint prior on the pairs $(\bfbet_{i,0}, \bfbet_{i,1})$ that explicitly accounts for the possibility that $\bfbet_{i,0} = \bfbet_{i,1}$.  More specifically, we introduce
a set of binary indicators $\zeta_1, \ldots, \zeta_I$ such that $\zeta_i = 1$ if and only if $\bfbet_{i,0} = \bfbet_{i,1}$, i.e., legislator $i$ exhibits the same ideal point on both final passage and procedural votes, and $\zeta_i=0$ otherwise.  We call legislators for which $\zeta_i = 1$ \textit{bridges}. Then, conditionally on $\zeta_i \in \{0, 1\}$ we let 
\begin{align}\label{eq:priorbeta}
  (\bfbet_{i,0},\bfbet_{i,1}) \mid \bfrho_{\beta}, \bfSig_{\beta}, \zeta_i & \simiid \begin{cases}
     \normal\left(\bfbet_{i,0} \mid \bfrho_{\beta}, \bfSig_{\beta} \right) \, \delta_{\bfbet_{i,0}}\left(\bfbet_{i,1}\right) & \zeta_i=1 , \\
     \normal\left(\bfbet_{i,0} \mid\bfrho_{\beta}, \bfSig_{\beta}\right)\normal\left(\bfbet_{i,1}\mid\bfrho_{\beta}, \bfSig_{\beta}\right) & \zeta_i=0 .\\
\end{cases}
\end{align}

Note that, when $\zeta_i = 0$ and therefore $\bfbet_{i,0} \ne \bfbet_{i,1}$, the two  ideal points $\bfbet_{i,0}$ and $\bfbet_{i,1}$ are given independent (albeit identical) priors.  Using the same values of $\bfrho_{\beta}$ and $\bfSig_{\beta}$ for both domains  reflects the idea that all these ideal points live in the same latent policy space. The hyperparameters $\bfrho_{\beta}$ and $\bfSig_{\beta}$ are learned from the data and given conditionally conjugate multivariate normal and inverse Wishart hyperpriors.

When reliable prior information about the identity of the bridge legislators is available, then the indicators $\zeta_1 , \ldots, \zeta_I$ can be treated as known covariates.  Furthermore, as long as at $d+1$ bridges are available, the two latent scales (for procedural and for final passage votes) are comparable, addressing the remaining identifiability issues that anchoring was not able to address on its own.  As we indicated before, this strategy is at the core of \textcite{shor2010bridge}.  Instead, in this paper we treat the indicators $\zeta_1 , \ldots, \zeta_I$ as unknown and devise a hierarchical prior that allows us to learn the identity of the bridge legislators and, more importantly for our goals, identify explanatory variables that are associated with revealing identical preferences on both voting domains.

\subsection{Explaining differences in ideal points}

Previous approaches to identifying bridge legislators from data (e.g., \citealp{Lof16_u} and \citealp{moser2021multiple}) have used exchangeable priors on the vector $\bfzeta = (\zeta_1, \ldots, \zeta_I)$ to control for multiplicities \citep{scott2006exploration,scott2010bayes}.  
Instead, this paper considers a hierarchical prior that allows us to introduce explanatory variables and formally test whether they are associated with the probability that a particular legislator is a bridge.  More specifically, let $\bfX$ be an $I \times p$ design matrix of (centered) explanatory variables, with $\bfx_i^T$ denoting the $i$-th row of $\bfX$, which in turn corresponds to the vector of observed explanatory variables associated with legislator $i=1,\ldots,I$.  We model the individual $\zeta_i$s using a logistic regression of the form
\begin{align}\label{eq:zetalik}
\zeta_i \mid \bfx_i, \eta_0, \bfeta \sim \Ber\left( \zeta_i \, \Bigg | \, \frac{1}{1 + \exp\left\{ - \left( \eta_0 + 
\bfx_i^T \bfeta \right)\right\}} \right) ,    
\end{align}
where $\eta_0$ is an intercept and $\bfeta = (\eta_1, \ldots, \eta_p)^T$ is a $p$-dimensional vector of unknown regression coefficients. 

In order to identify explanatory variables that affect the probability that a particular legislator is a bridge, we adopt a prior for $\bfeta$ that allows for coefficients to be exactly zero. With this goal in mind, let $\bfxi = (\xi_1, \ldots, \xi_p)$ be a binary $p$-dimensional vector such that $\xi_k = 1$ if variable $k$ is included in the model (i.e., if $\eta_k \ne 0$) and $0$ otherwise (i.e., if $\eta_k = 0$). Also, let $\bfX_{\bfxi}$ be the matrix created by retaining the columns of $\bfX$ for which the corresponding entries of $\bfxi$ are equal to 1 and, similarly, let $\bfeta_{\bfxi}$ be the vector made of the entries of $\bfeta$ for which the corresponding entries of $\bfxi$ are equal to 1.  Then, conditional on $\bfxi$, 
the prior for $\bfeta$ takes the form 
\begin{align*}
f(\bfeta \mid \bfxi, \bfX) = f(\bfeta_{\bfxi} \mid {\bfxi}, 
\bfX_{\bfxi}) \prod_{\{ k : \xi_k = 0\}} \delta_{0} (\eta_k) ,
\end{align*}
where $\delta_0(\cdot)$ again denotes a point mass at zero, and $f(\bfeta_{\bfxi})$ corresponds to a g-prior of the form 
\begin{align}\label{eq:gprior}
\eta_{\bfxi} \mid g, \bfxi, \bfX_{\bfxi} \sim N\left(\eta_{\bfxi} \, \Big | \, 0, 4 g \left(\bfX_{\bfxi}^T \bfX_{\bfxi} \right)^{-1}\right) .
\end{align}
In the sequel, we work with $g=I$, which means that \eqref{eq:gprior} can be interpreted as an (approximately) unit information prior under an imaginary training sample (see \citealp{sabanes2011hyper} for additional details). 

Finally, we follow \textcite{scott2010bayes} and use a Beta-Binomial prior for $\bfxi$,
\begin{align}\label{eq:modelprior}
    f(\bfxi) = \frac{\Gamma\left(p^*_{\bfxi} + 1 \right)\Gamma\left(p - p^*_{\bfxi} + 1\right)}{\Gamma( p + 2)} = \int_0^1 \upsilon^{p^*_{\bfxi}} (1 - \upsilon)^{p - p^*_{\bfxi}} d \upsilon,
\end{align}
where $p^*_{\bfxi}=\sum_{k=1}^p \xi_k$ is the size of the model.
This prior has several advantages.   One of them is interpretability.  As the last equality in \eqref{eq:modelprior} shows, it can be interpreted as the result of assuming a common (but unknown) prior probability of inclusion $\upsilon$ for every variable in the model, and then assigning $\upsilon$ a uniform distribution.  Hence, it implies that the marginal probability of inclusion for each variable is $1/2$.  However, because the entries of $\bfxi$ are correlated, the prior induces multiplicity control by assigning a uniform distribution on the size of the model $p^*_{\bfxi}$ (see \citealp{scott2010bayes} for additional details).

The model is completed by specifying the prior for the intercept $\eta_0$.  Because of the hierarchical nature of our model, we avoid the use of improper flat priors and instead adopt a standard logistic prior with density
$$
f(\eta_0) = \frac{\exp\left\{ \eta_0 \right\}}{ \left( 1 + \exp\left\{ \eta_0 \right\} \right)^2}.
$$
Employing a prior for $\eta_0$ that is independent of the prior on $\bfeta_{\bfxi}$ is natural in this case because we work with a centered design matrix $\bfX$.  The choice of a logistic prior implies that, under the null model where none of the variables affect the bridging probabilities, the implied prior on $\Pr(\zeta_i = 1)$ 
reduces to a uniform distribution on the $[0,1]$ interval.  Hence, under the null model, this approach is equivalent to that of \citet{Lof16_u}.  Another important feature of this prior is that it typically leads to stable estimates in the presence of separation (e.g., see \citealp{boonstra2019default}).  This is important because separation is a potential concern in our setting (particularly for years in which the the number of bridges is very high), and because the latent nature of $\bfzeta_i$ makes it impossible to check for the presence of separation before fitting our model.\footnote{See \textcite{rainey2016dealing} for a more detailed discussion of separation issues in logistic regression.}

\section{Computation}\label{sec:computation}

Posterior inference is carried out using a Markov chain Monte Carlo (MCMC) algorithm.  Our approach relies heavily on the data augmentation approach introduced in \textcite{polson2013bayesian}.  More specifically, we rely on the fact that the Bernoulli likelihood can be written as a mixture of the form
\begin{align*}
    \frac{\exp\left\{ z_i \psi_i \right\}}
    {1+\exp{\left\{\psi_i\right\}}}
    & \propto\exp\left\{(z_i - 1/2) \psi_i \right\}\int_0^\infty \exp \left\{- \frac{\omega_i}{2} \psi_i^2 \right\}f_{PG}(\omega_i \mid 1,0) d\omega_i ,
\end{align*}
where $f_{PG}(\omega \mid a,b)$ denotes the density of a P\`olya-Gamma random variable with parameters $a$ and $b$. 
We use this data augmentation twice in our algorithm:  once for the likelihood of the observed data in \eqref{eq:likelihood2}, and then again for the distribution of the bridge indicators in  \eqref{eq:zetalik}.

Introducing P\`olya-Gamma auxiliary variables dramatically simplifies the formulation of our computational algorithm.  As a consequence, most of the full conditionals of interest can be sampled directly (i.e., most of the steps of our MCMC algorithm reduce to Gibbs sampling steps).  The only two exceptions are the full conditional posterior distribution for $\eta_0$, which does not belong to a known family, and the full conditional posterior distribution for $(\bfxi, \bfeta)$, which is a mixture with an unmanageable number of components.  For $\eta_0$, we  develop a Metropolis-Hastings algorithm with heavy-tailed independent proposals (as opposed to a random-walk Metropolis Hasting) that has a very high acceptance rate.  For $(\bfxi,\bfeta)$, we use a random walk Metropolis-Hasting proposal for the marginal full conditional of $\bfxi$ obtained after integrating out $\bfeta$, and then sample $\bfeta$ directly from its Gaussian full conditional distribution. 
For additional details, please see the implementation available at \url{https://github.com/e-lipman/ProcFinalUSHouse}.

We use the empirical averages of the posterior samples to approximate posterior summaries of interest.  For example, we summarize the posterior distribution on models through posterior  inclusion probabilities (PIPs) associated with each variable in the design matrix $\bfX$.  The PIP for variable $k$ is defined as
\begin{equation}\label{eq:PIP}
\Pr(\xi_k=1 \mid \mbox{data})=\E\left\{ \xi_k \mid \mbox{data} \right\} \approx \frac{1}{S} \sum_{s=1}^{S}  \xi_k^{(s)} ,
\end{equation}
where $S$ represents the total number of samples drawn from the posterior distribution and $\xi_k^{(s)}$ denotes the $s$-th sampled value for the parameter $\xi_k$.  PIPs provide a measure of the uncertainty associated with the influence of any given variable on the probability that a legislator is a bridge.  Point and interval estimates of other quantities of interest can be obtained in a similar way.

\subsection{Parameter identifiability}

The $\mu_i$s, $\alpha_j$s, $\beta_{i,0}$s and $\beta_{i,1}$s are not identifiable.  Hence, if we are interested in performing inferences on them, we need to be careful how the samples from the Markov chain Monte Carlo algorithm are used to construct the empirical averages that serve as posterior summaries.

In this manuscript, identifiability constraints are enforced after each iteration of the Markov chain Monte Carlo algorithm rather than through constrained priors. This approach is often referred to in the literature as parameter expansion \citep{liu1999parameter,schliep2015data}.  One such constraint refers to the need to have enough bridges to make the scale associated with the two voting domains comparable.  In our application we consider a one-dimensional latent space (i.e., $d=1$).  Hence, we enforce  $\zeta_{\cdot} = \sum_{i=1}^{I} \zeta_i \ge d+1 = 2$.  In practice, the posterior distributions for $\zeta_{\cdot}$ puts all of its mass on much larger values for all the datasets discussed in Section \ref{sec:results}, which means that the constraint is never binding and no adjustment to the posterior samples was required in any of the datasets we analyzed.  A second identifiability constraint requires that both scales should point in the ``same direction'' (e.g., Republican legislators should tend to have positive values on both scales).  This can be easily accomplished by enforcing a common \text{sign} (but not necessarily a common \text{value}) for the two ideal points (procedural and final passage) associated with one carefully selected legislator (in our case, the Republican party leader).  A third constraint is  associated with identifying the absolute scale, and is enforced by translating and scaling the ideal points of the legislators so that $d+1$ of them (the anchors) are fixed to specific values in one of the voting domains 


\section{Example Application: 
 Procedural and final passage Voting in the U.S.\ House of Representatives, 1973-2014} \label{se:example}

In this section we apply the model we just described to voting in the U.S.\ House of Representatives in two domains: procedural motions and final passage motions.  Procedural votes are broadly viewed separately from substantive votes \citep{ther06-party}. The determinants of legislative voting on procedural motions in the House of Representatives include party affiliation and the presence of other votes on the same day \citep{patt10-dilatory}. Party matters for procedural motions, but there are differences between the majority party and the minority party \citep{cars14-procedural}.  On the other hand, the determinants of legislators' voting on final passage motions include electoral concerns and the possibility of receiving political appointments \citep{good07-rollcall}. In the specific case of the U.S.\ House of Representatives, legislative voting on final passage motions is influenced by factors such as the importance of electoral considerations \citep{good04-lameduck}, 
access to pork benefits \citep{shin17-choice}, and cross-pressure pitting the preferences of legislators' constituency against those of their party leaders \citep{kirk14-partisanship}.

Scholars have long been interested in the influences on legislators' voting both in the U.S. Congress, which is where our empirical example occurs, as well in other parliaments and legislatures. We focus on three classes of explanations. Broadly, scholars have studied three classes of covariates when explaining legislative voting behavior: \emph{constituency-level characteristics} (how conservative/ liberal is a member's district? Are there military bases located in their district? How much agricultural activity takes place?); \emph{legislator-level characteristics} (is the member male or female? Is the member from a `safe district?') and; what we call \emph{chamber-level characteristics} (is the member in the majority party? Was electronic voting used in the chamber or not? Under what rule was the roll call taken?).\footnote{This is by no means exhaustive. Other scholars have   looked at the effect of lame duck sessions on legislative voting, effective term limits, etc.}

In general, legislators must balance the competing interests of their constituents, their political parties, interest groups, and their own ideological beliefs. Constituent interests play a significant role in legislative voting \citep{lope07-strategic}. Political parties also strongly influence how legislators vote \citep{hix16-governmentopposition}. Public interest lobbies, ideology, and PAC contributions also predict legislative voting \citep{kau13-congressman}. Constituency matters \citep{flei04-shrinking,bond02-disappearance}, while gender does not seem to have an effect on legislative voting \citep{schw04-gender}. We return to these general findings when discussing our results (Section \ref{sec:results}).


\subsection{Data}\label{se:data}

This section describes the data that serves as the basis for our illustration and provides a brief descriptive analysis.
Similar to \citet{jess14-two}, we consider the roll-call voting records in the U.S.\ House of Representatives starting with the 93\textsuperscript{rd} Congress (in session between January 1973 and January 1975).  This coincides with the introduction of electronic voting in the House, which led to a marked increase in the number of roll-call votes, as well as a change in their nature \citep{jessee2014two}.  Our analysis extends to the 113\textsuperscript{th} Congress, which is as far as our data sources on potential covariates extend (please see below).  

\begin{figure}[!ht]
\centering
\includegraphics[width=.47\textwidth]{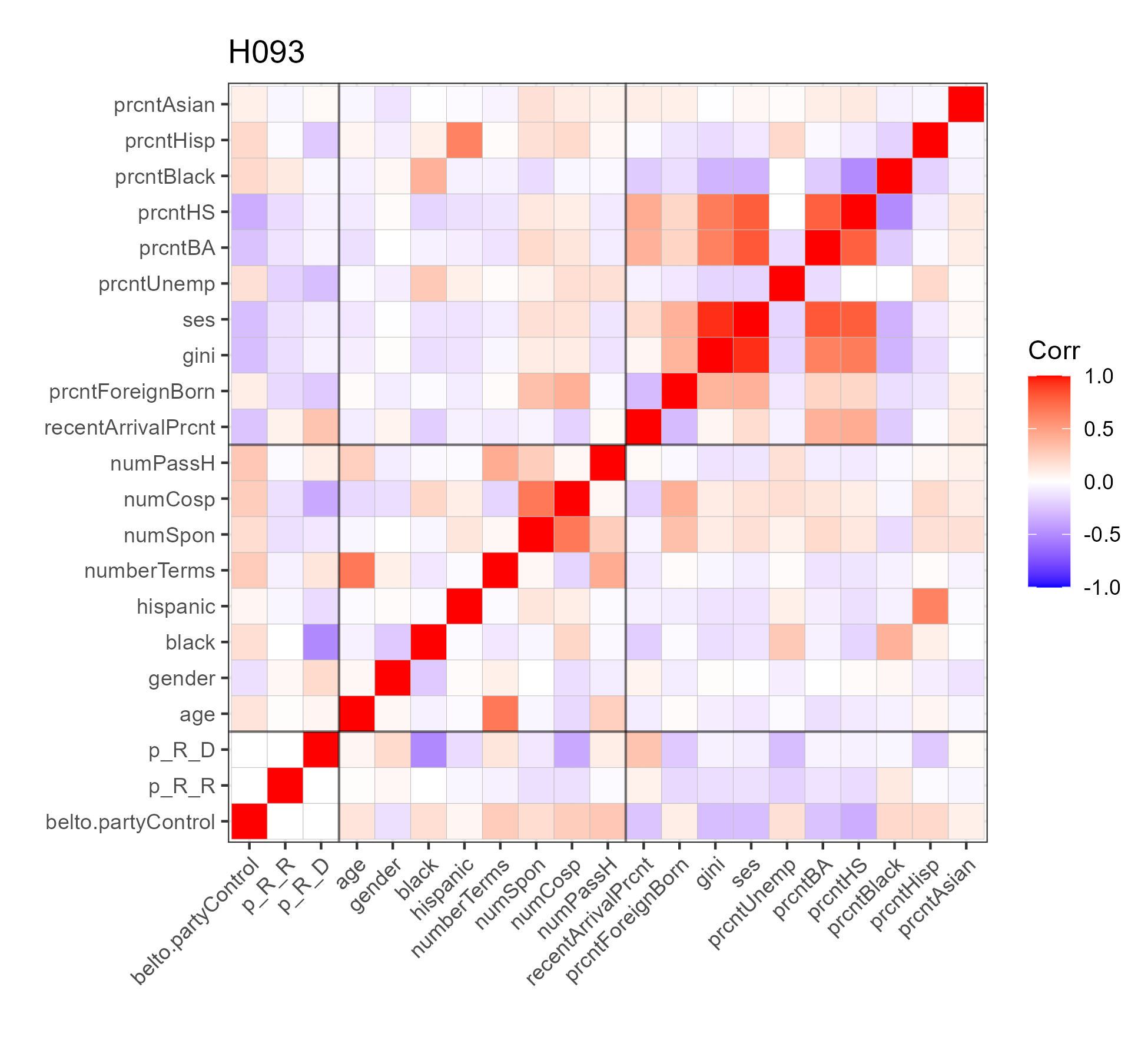}
\includegraphics[width=.47\textwidth]{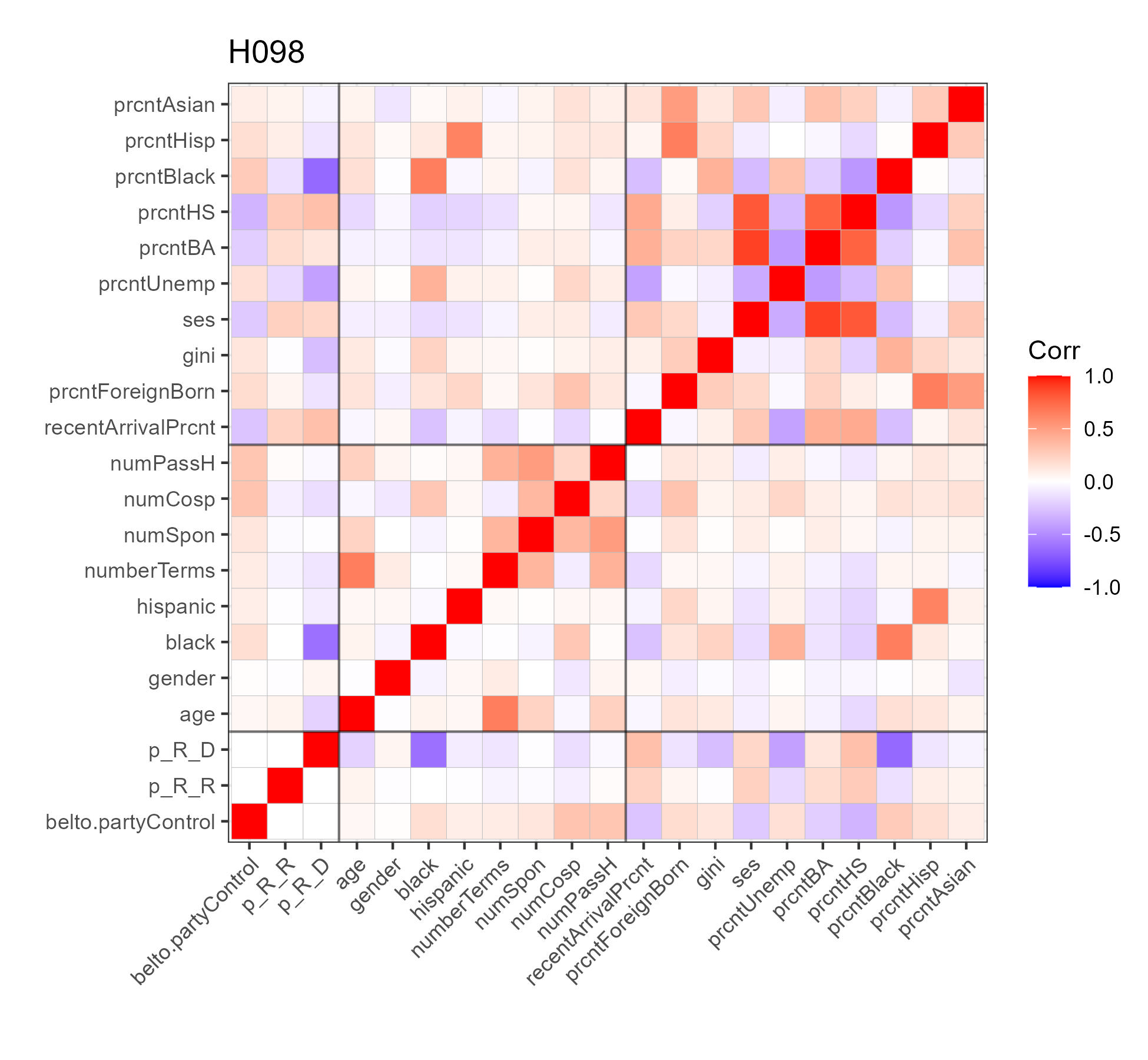}
\includegraphics[width=.47\textwidth]{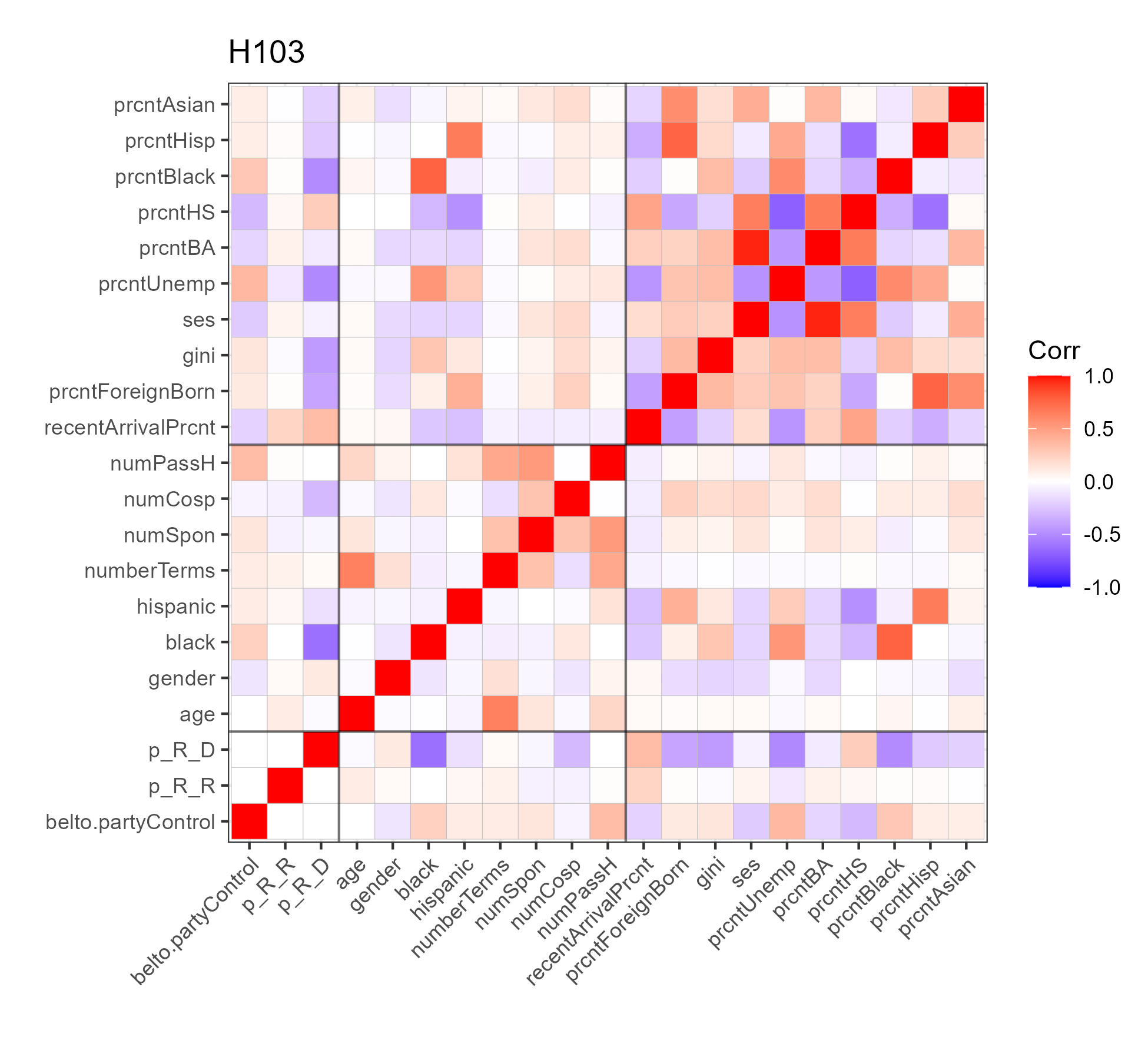}
\includegraphics[width=.47\textwidth]{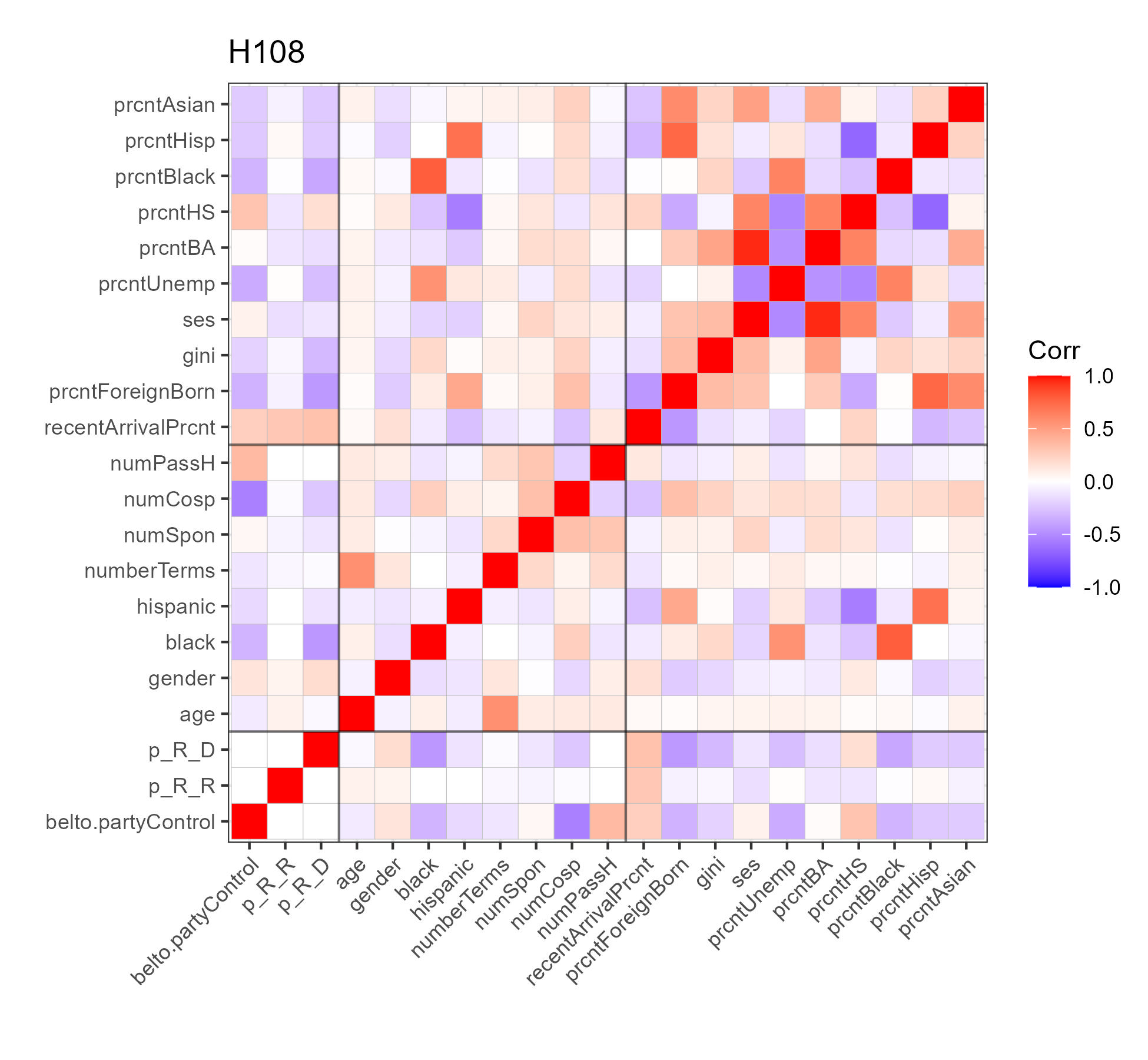}
\caption{Correlation plots for the regression matrices associated with four Houses under study: 
 the 93\textsuperscript{rd}, 98\textsuperscript{th}, 103\textsuperscript{rd} and 108\textsuperscript{th} Houses. }\label{fig:cormatrices}
\end{figure}

In our analysis, we consider 21 variables as potentially explaining the tendency of individual legislators to reveal identical preferences across both procedural and final passage votes.\footnote{See  \ref{ap:variable} for details.}  The constituency-level covariates include measures of racial diversity and economic inequality, as well as the political leaning of the constituency. Political leaning is measured by the percentage of the district's vote that went to the Republican candidate in the most recent presidential election.  The legislator-level covariates include age, gender, and race, as well as measures of activity while in office, such as the number of sponsored bills. The chamber-level covariates include whether the legislator's party was in the majority. 
%
Figure \ref{fig:cormatrices} presents the correlation matrices for the covariates associated with four different Houses.  Overall, correlations seem to be low, and the highest appear to be among constituency-related covariates.

Two variables that will appear later to be particularly important in explaining differences in voting behavior between procedural and final passage votes are (1) the percentage of the two-party presidential vote share received by the Republican presidential candidate in the most recent presidential election, and (2) the party affiliation of the legislator and, in particular, whether their party was in control of the House.  We proceed with a brief descriptive analysis of each of these variables.  

\begin{figure}[!t]
\centering
\includegraphics[width=0.85\textwidth]{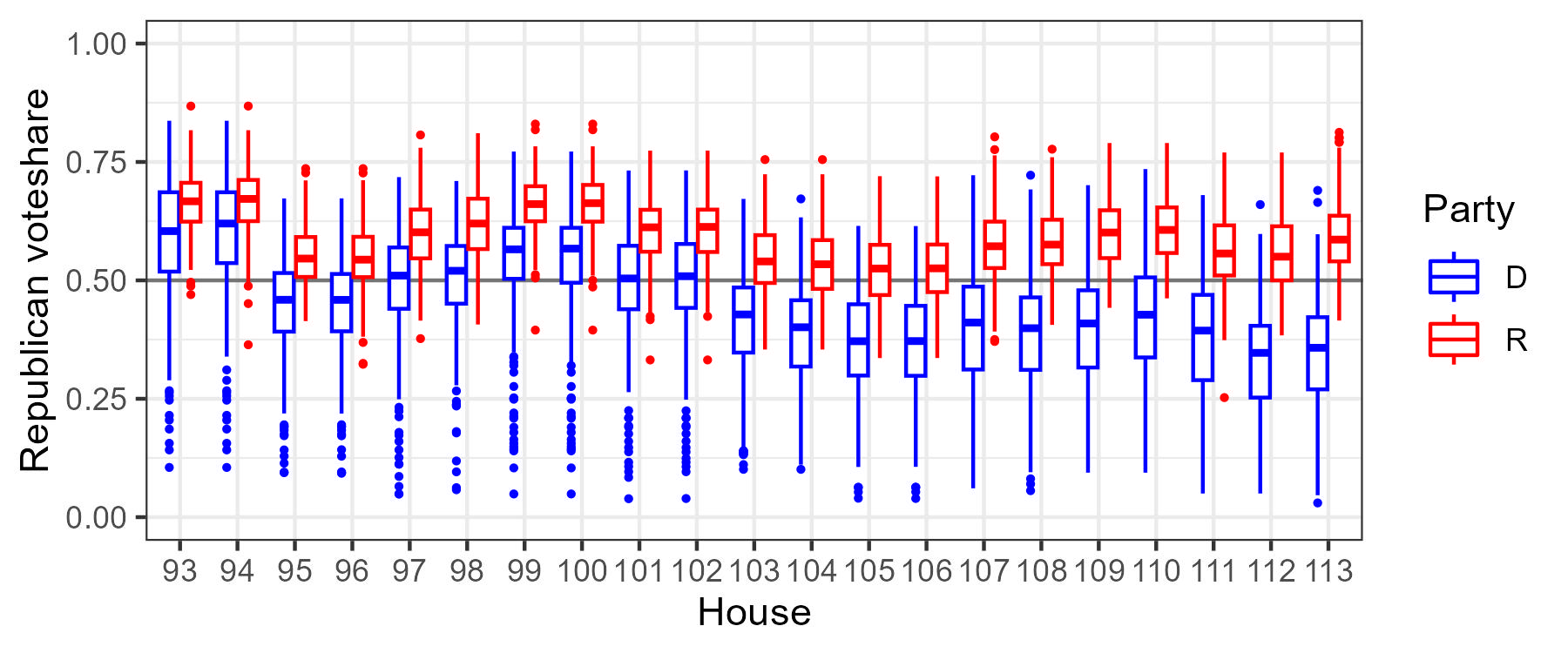}
\caption{Percentage of the two-party presidential vote share received by the Republican presidential candidate in the legislator's district in the most recent presidential election.  Districts are divided into Republican and Democrat depending on the party affiliation of the legislator representing that district during the corresponding House.}
\label{fig:consistuancy}
\end{figure}

Figure \ref{fig:consistuancy} shows boxplots of the Republican vote share for each House, broken down by the party of affiliation of the legislator representing the districts. Earlier in the time series, the median for both parties is often at or above 50\%, indicating that many representatives affiliated with the Democratic party represent districts that are either toss-ups or lean Republican.  Furthermore, in these early Congresses, the boxes for the two parties tend to overlap.  On the other hand, towards the second part of the time series, we can see districts becoming more polarized, with very few legislators representing districts where the majority of their constituency voted for the opposite party during the most recent presidential election.  A second interesting feature of these data is that the range of districts held by Democrats tends to be much larger than that of the districts held by Republicans.  In particular, there are many districts that lean heavily Democratic on presidential elections (with Democratic vote share often above 90\%), but there are few or no districts that lean as heavily Republican.  The presence of these distinct patterns motivated us to include in our analysis the interaction between the constituency ideological leaning (as represented by the percentage of Republican vote in presidential elections) and the party of affiliation of the legislator. 

\begin{figure}[!ht]
\centering
\includegraphics[width=0.85\textwidth]{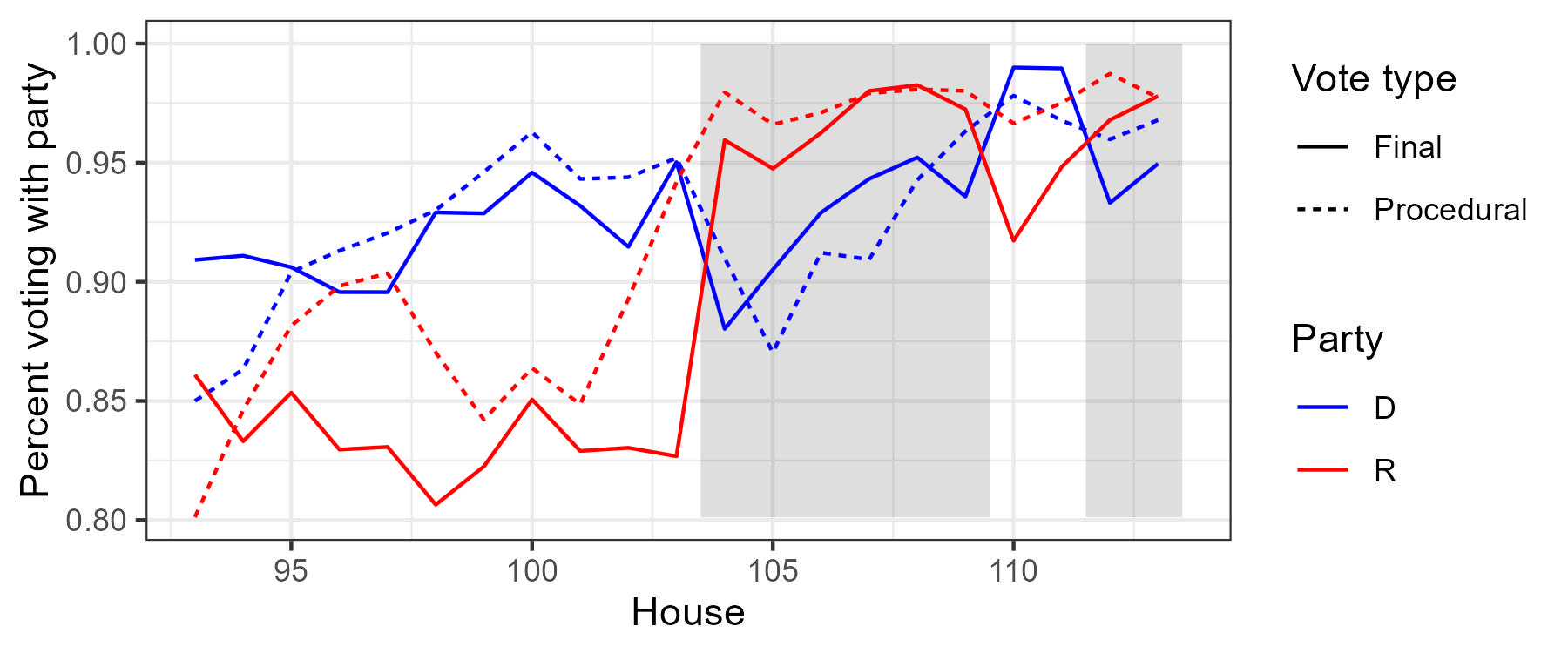}
\caption{Percentage of legislators in each party voting with the party majority. Shown separately for procedural and final passage votes. Shaded regions represent sessions with a
Republican majority and unshaded regions represent sessions with a Democratic majority.
}\label{fig:with_party}
\end{figure}

In order to explore the role that party affiliation might play in voting behavior on procedural and final passage votes, Figure \ref{fig:with_party} presents the median (across legislators) of the proportion of times that each legislator voted with their party, broken down by vote type and party.  We can see that the value has been historically high (generally over 80\%), and that there is also evidence of an increasing trend over time.  This increase is particularly dramatic for the Republican party during the 104\textsuperscript{th} House, when the percentage jumps by about 10\% for both types of votes.  This observation is consistent with previous literature pointing out that party influence on legislator's voting behaviors has been variable over time (e.g., see \citealp{aldrich1995parties} and \citealp{sinclair2006party}).  We also observe that the percentage is often (but not always) slightly higher for procedural than for final passage votes.  Again, this observation is consistent with previous literature suggesting that party influence in the U.S.\ House of Representatives tends to be higher on procedural votes (e.g., see \citealp{roberts2003procedural},   \citealp{roberts2007statistical}, \citealp{lee2009beyond} and \citealp{gray2022messaging})\footnote{Whether this is also true in the Senate is debated, e.g., see \textcite{algara2019member} and \textcite{smith2014senate}.}. 
Finally, we note that vote cohesion has tended to be higher (for both vote types) for the party in control of the House.

\subsection{Results} \label{sec:results}

This section reports the results obtained by applying the model described in Section~\ref{sec:methods} to the data just described. 
We verify convergence of our algorithm by running four chains for each dataset, started at over-dispersed initial values.  For each chain, we monitor the trace plot and the Gelman-Rubin statistic  \citep{GeRu92,vats2021revisiting} of the (unnormalized) joint posterior distribution of all parameters, the total number of bridge legislators, $\zeta_{\cdot} = \sum_{i=1}^{I} \zeta_i$ (as well as the number of bridge legislators broken down by party affiliation), the value of the linear predictor $\eta_0 + \bfx_i^T\bfeta$ for 
each legislator, and the number of predictors in the model, $p^{*}_{\xi}$.  Most of the results in Section \ref{sec:results} are based on 60,000 posterior samples for each House, obtained by combining four independent runs of the Markov chain Monte Carlo algorithm, each consisting of 15,000 samples obtained after a burn-in period of 20,000 samples and thinning every 20 samples.  The one exception is the 105\textsuperscript{th} House.  Unlike the other Houses under study, the posterior distribution of the model size $p^{*}_{\bfxi}$ appears to be bimodal in this case.  The dominant mode favors relatively small models that include between 2 and 5 variables, while the second mode favors large models that can have up to 15 explanatory variables.  These two modes also manifest themselves in the posterior distribution of $\bfzeta$, with the dominant mode supporting what appears to be a superset of the bridges supported by the lower-probability one.  To ensure that both modes are explored thoroughly and that we can accurately estimate their relative importance, we base our inference in this House on eight chains, each of which consists of 20,000 iterations obtained after thinning every 25 samples and a burn-in period of 20,000 iterations.  

As we mentioned before, we work with a one dimensional policy space, i.e., $d=1$.  We assign $\rho_{\mu}$ and $\rho_{\beta}$ standard Gaussian priors, and $\kappa^2_{\mu}$, $\kappa^2_{\alpha}$ and $\Sigma_{\beta}^2$ inverse Gamma priors with shape parameters 2 and scale parameter 1 (so that they have mean 1 and infinite variance).  This choice is consistent with the logic underlying the logistic scale and with the recommendations in \textcite{bafumi13andrew}. Furthermore, we assign $\omega_{\alpha}$ a uniform distribution.

To investigate the sensitivity of the model to prior choices, we consider a few alternative prior specifications for the key regression parameters $\eta_0$ and $\bfeta_{\bfxi}$.  In particular, we explored replacing the logistic prior for $\eta_0$ with a standard Gaussian prior and replacing the g-prior on $\bfeta$ with a mixture of g-priors that relies on the robust prior developed in \textcite{bayarri2012criteria}.  The results were quite robust to these changes, although we do note that the mixture of g-priors tended to favor slightly lower inclusion probabilities for most of the variables.

\subsubsection{Aggregate analysis}

\begin{figure}[t]
\centering
\includegraphics[width=0.85\textwidth]{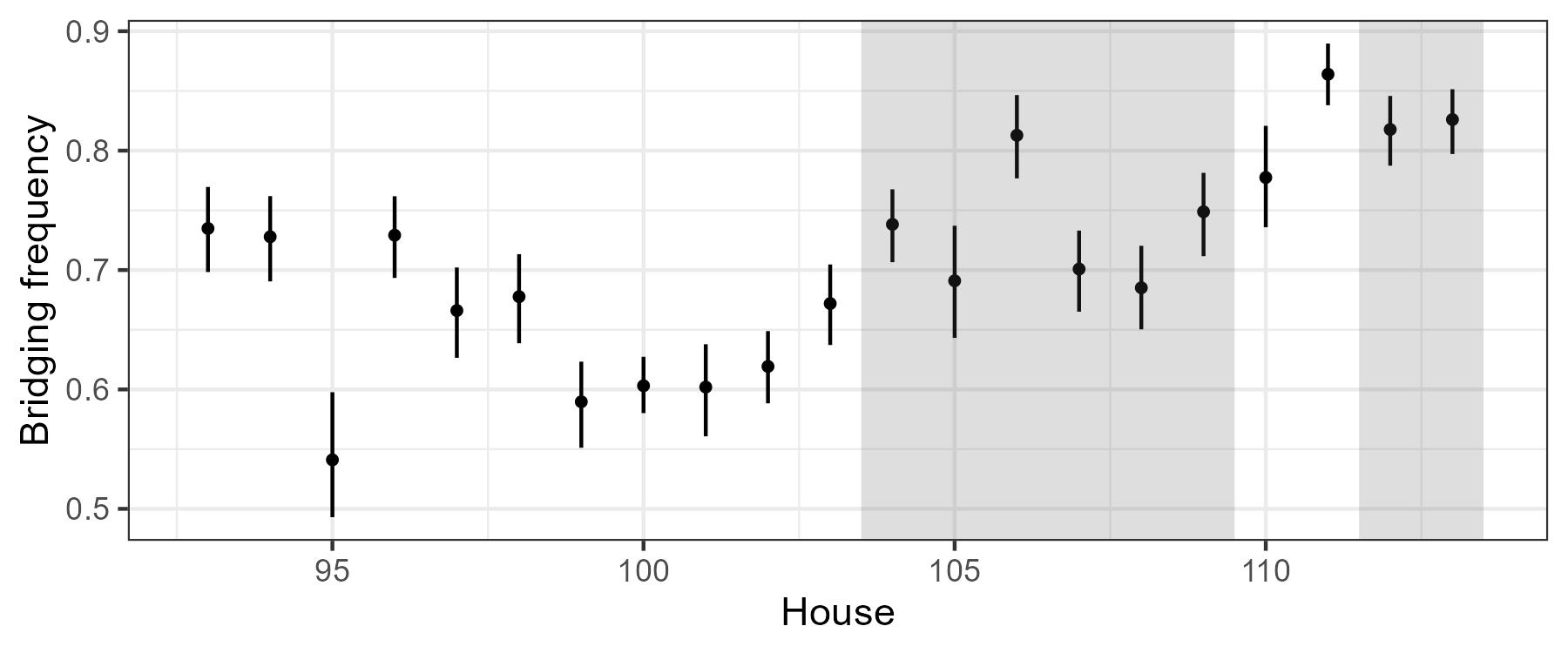}
\caption{Posterior mean and 95 percent credible intervals for the bridging frequency in the U.S.\ House of Representatives from the 93\textsuperscript{rd} through 113\textsuperscript{th} sessions of congress. Shaded regions represent sessions with a Republican majority and unshaded regions represent sessions with a Democratic majority.}
\label{fig:ASF}
\end{figure}

We begin with a chamber-level analysis. Figure~\ref{fig:ASF} shows the posterior mean and 95 percent credible intervals for the Bridging Frequency (BF) for each House.  The BF is defined as the proportion of bridges, i.e., legislators in each House whose revealed preferences are the same across both procedural and final passage votes.  From equation \eqref{eq:priorbeta} we have:
\begin{equation}\label{eq:BF}
BF = \frac{1}{I}\sum_{i=1}^I \zeta_i.
\end{equation}

Figure~\ref{fig:ASF} suggests a decline in bridging frequencies during the 70s and 80s, followed by a steady increase  starting perhaps with the 102\textsuperscript{nd} House (1991-1992).  The decrease in the number of legislators that vote differently on procedural and final passage votes has at least two potential explanations, both of which are related to an increased tendency for legislators to vote along party lines (recall Figure \ref{fig:with_party} and the associated discussion in Section \ref{se:data}).  One explanation, which is consistent with the arguments in \textcite{jess14-two}, is that party influence on final passage votes has increased since the early 1990s.  Under this interpretation, our results provide support for the ``procedural cartel'' theory of  political parties (e.g., see \citealp{cox2005setting} and \citealp{clark2012examining}). 
An alternative explanation is provided by constituency pressure.  Increasing polarization within congressional districts driven (among other factors) by the sorting of political identities along the urban/rural divide and by aggressive gerrymandering, has resulted in fewer competitive districts overall.  There is evidence that fewer competitive districts have meant fewer opportunities for moderate candidates \citep{barber2015causes,thomsen2014ideological,thomsen2017opting}.  This has, in turn, led to primary elections having an outsized impact on the final outcome of congressional elections \citep{kaufmann2003promise,abramowitz2008polarization,bafumi2010leapfrog,hall2018punishes}.  In this context, there are strong incentives for legislators to vote along party lines on all votes in order to avoid primary challenges, independently of any leadership pressure.
%

\subsubsection{Legislator-level analysis:  explaining the bridges}

\begin{figure}
\centering
\includegraphics[width=0.85\textwidth]{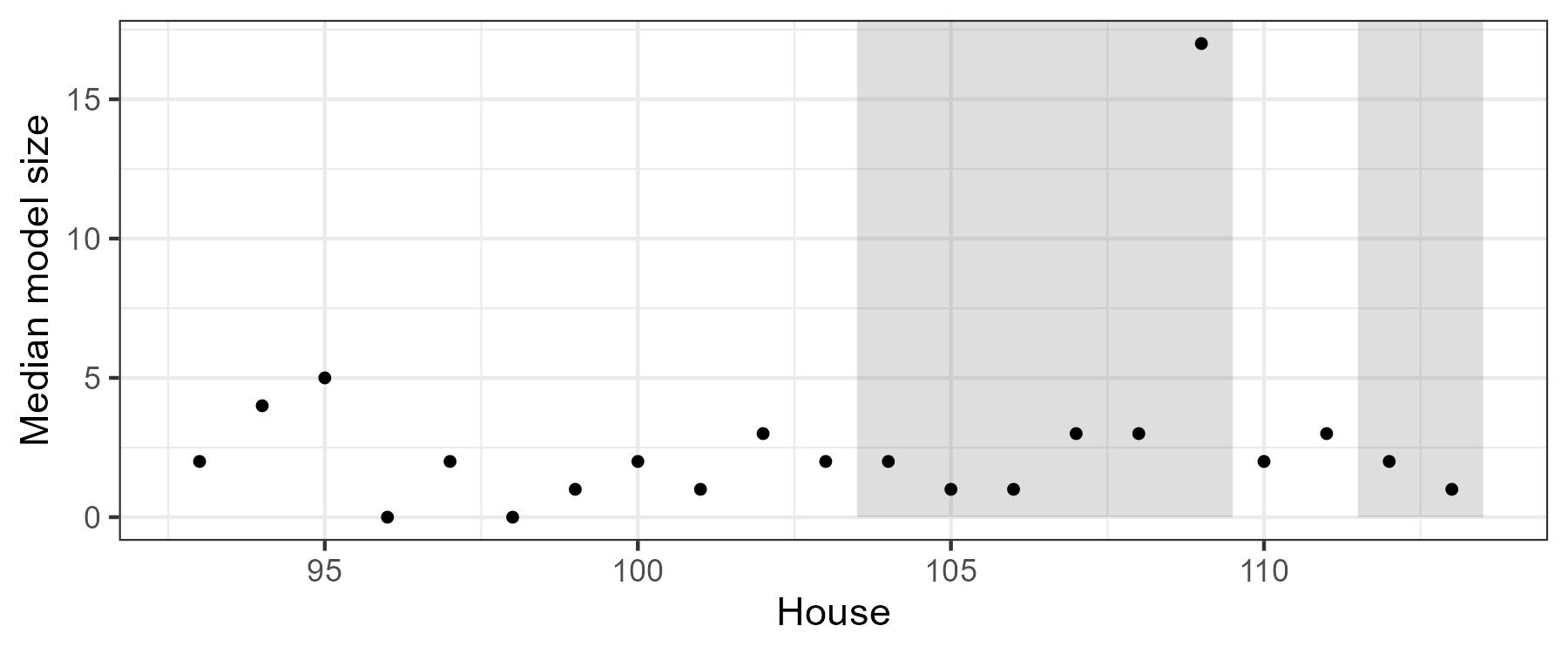}
\caption{Number of covariates included in the posterior median model for each House under consideration. Shading indicates congresses with a Republican majority.}
\label{fig:modelsize}
\end{figure}

We turn our attention to the identification of variables that are correlated with bridging behavior. Figure \ref{fig:modelsize} shows the number of variables with a PIP greater than 0.5 for each House (recall Equation \eqref{eq:PIP}).  In almost all cases, this number is between 0 and 5 variables.  The only outlier is the  109\textsuperscript{th} House, in which 17 variables appear to be significant.  This result was surprising.  While the 109\textsuperscript{th} House met for only 242 days (the fewest since World War II), neither the absolute nor the relative number of procedural and final passage votes appear to be particularly out of the ordinary.  One notorious fact about this House is the large number of political scandals that affected its members, and particularly those in the Republican party.  We speculate that the Republican majority leader Tom DeLay's campaign finance scandal and his eventual resignation from Congress (along with the  Bob Ney, Randy Cunningham and Mark Foley scandals) might have led to a breakdown in Republican party discipline in the House.  This hypothesis seems consistent with the procedural cartel theory we discussed in the previous section, and supports the idea that party discipline might be the main driver behind the higher bridging frequencies, and not the smaller number of competitive districts.

\begin{figure}
\centering
\includegraphics[width=0.85\textwidth]{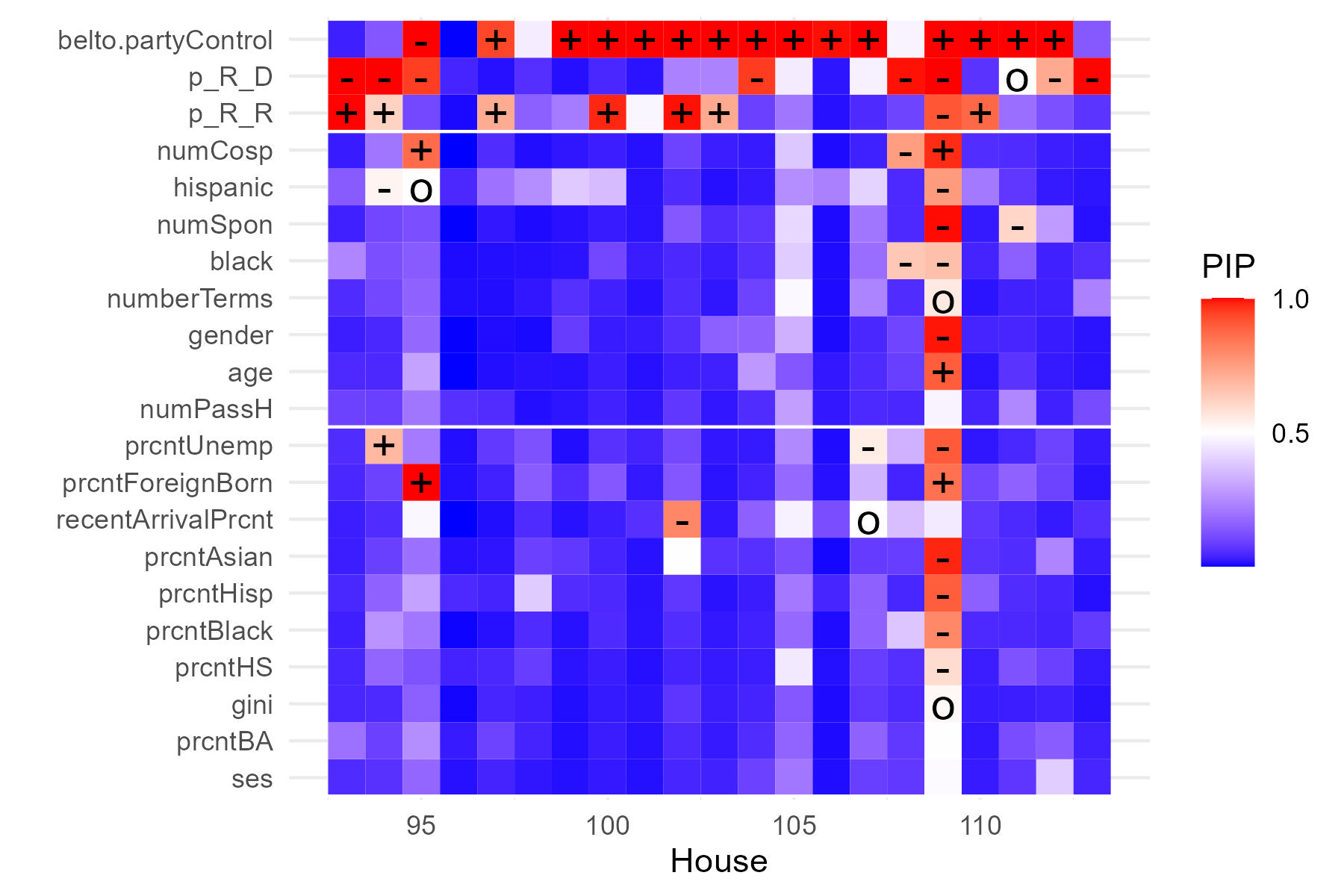}
\caption{Heatmap showing posterior inclusion probabilities (PIP) for each covariate in each House under study. Plus and minus signs indicate covariates with positive and negative posterior medians, respectively. Horizontal lines divide the covariates into groups: the three most commonly-selected covariates are in the top group, followed by legislator characteristics (middle), and the remaining constituency characteristics (bottom).}
\label{fig:PIPs}
\end{figure}

\begin{figure}[!ht]
  \begin{subfigure}[b]{\textwidth}
  \centering
  \includegraphics[width=0.8\textwidth]{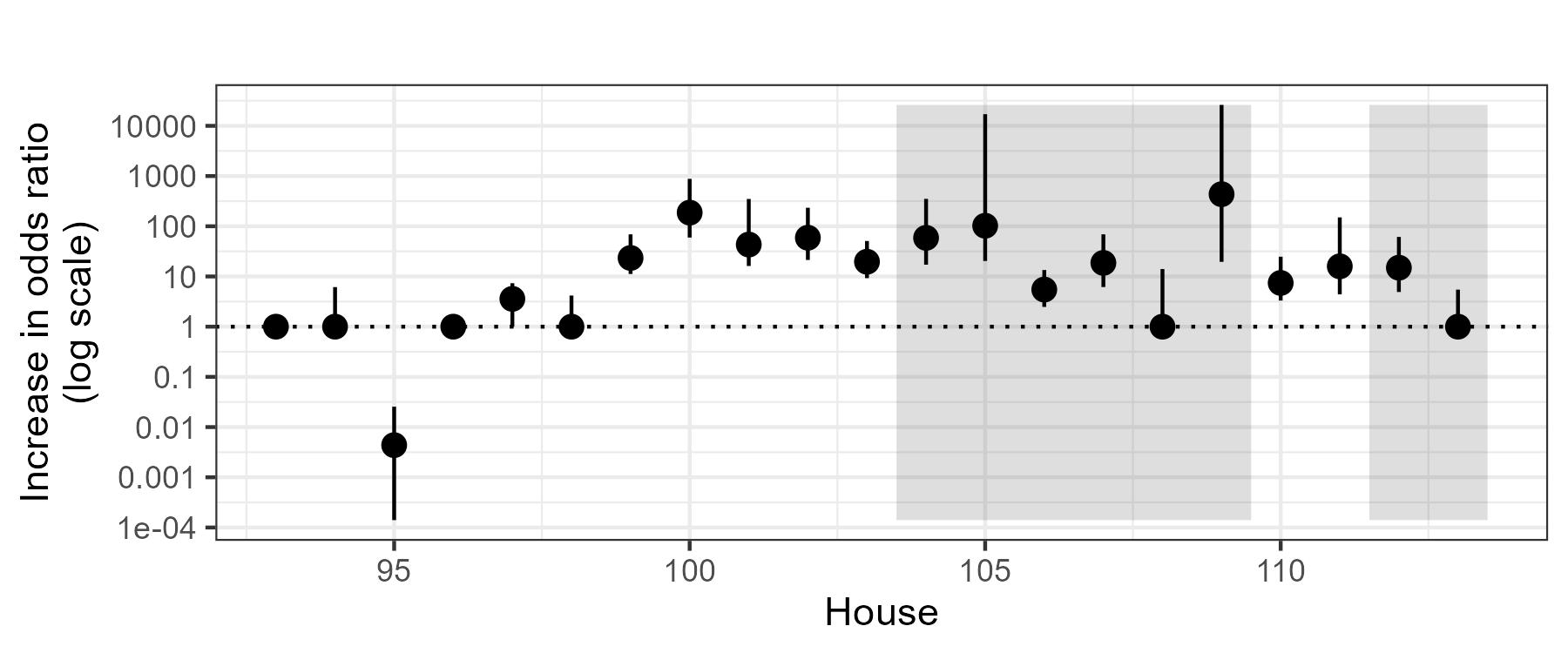}
  \caption{Influence of belonging to the majority party on odds of being a bridge legislator}\label{fig:beltoparty}
  \end{subfigure}
  \begin{subfigure}[b]{\textwidth}
  \centering
  \includegraphics[width=0.8\textwidth]{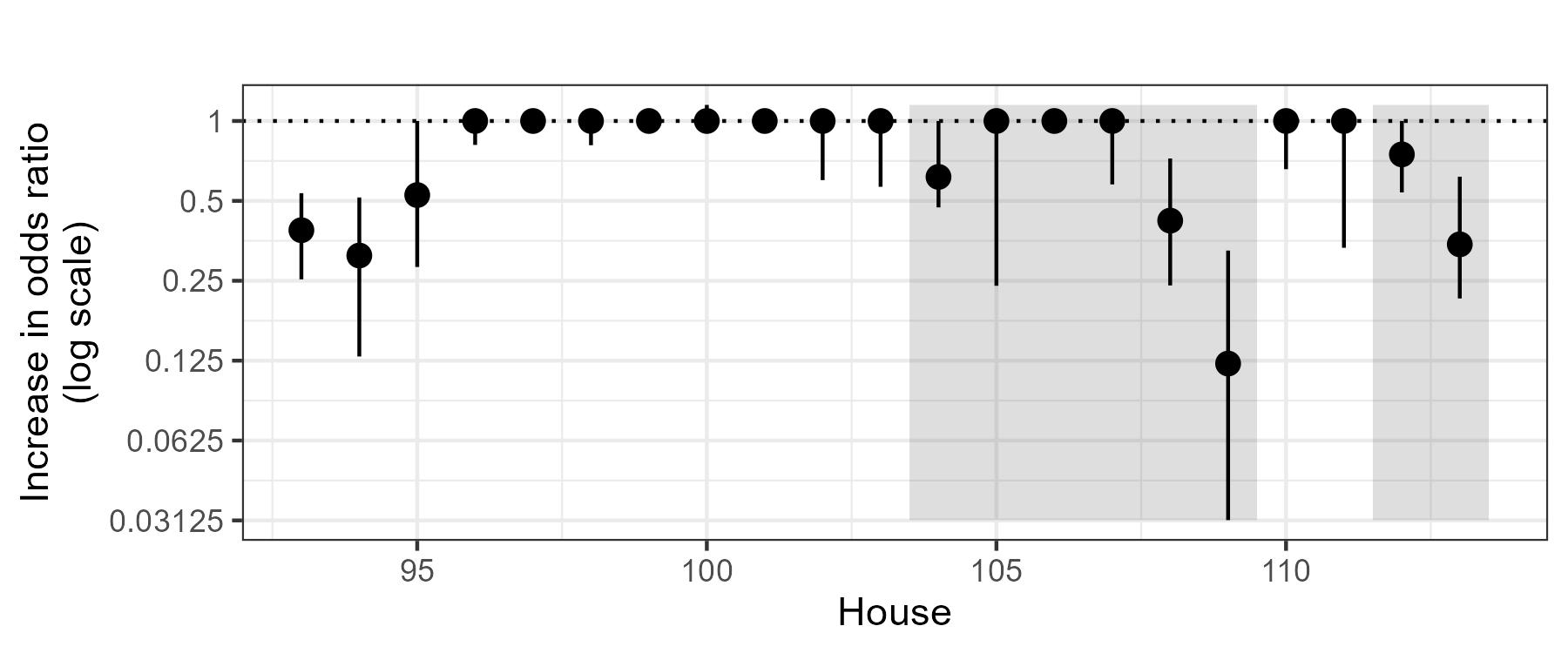}
  \caption{Influence of a $5\%$ increase in the proportion of constituency-level Republican vote in the most recent presidential election on odds of being a bridge legislator, for Democrats only}\label{fig:prD}
  \end{subfigure}
  \begin{subfigure}[b]{\textwidth}
  \centering
  \includegraphics[width=0.8\textwidth]{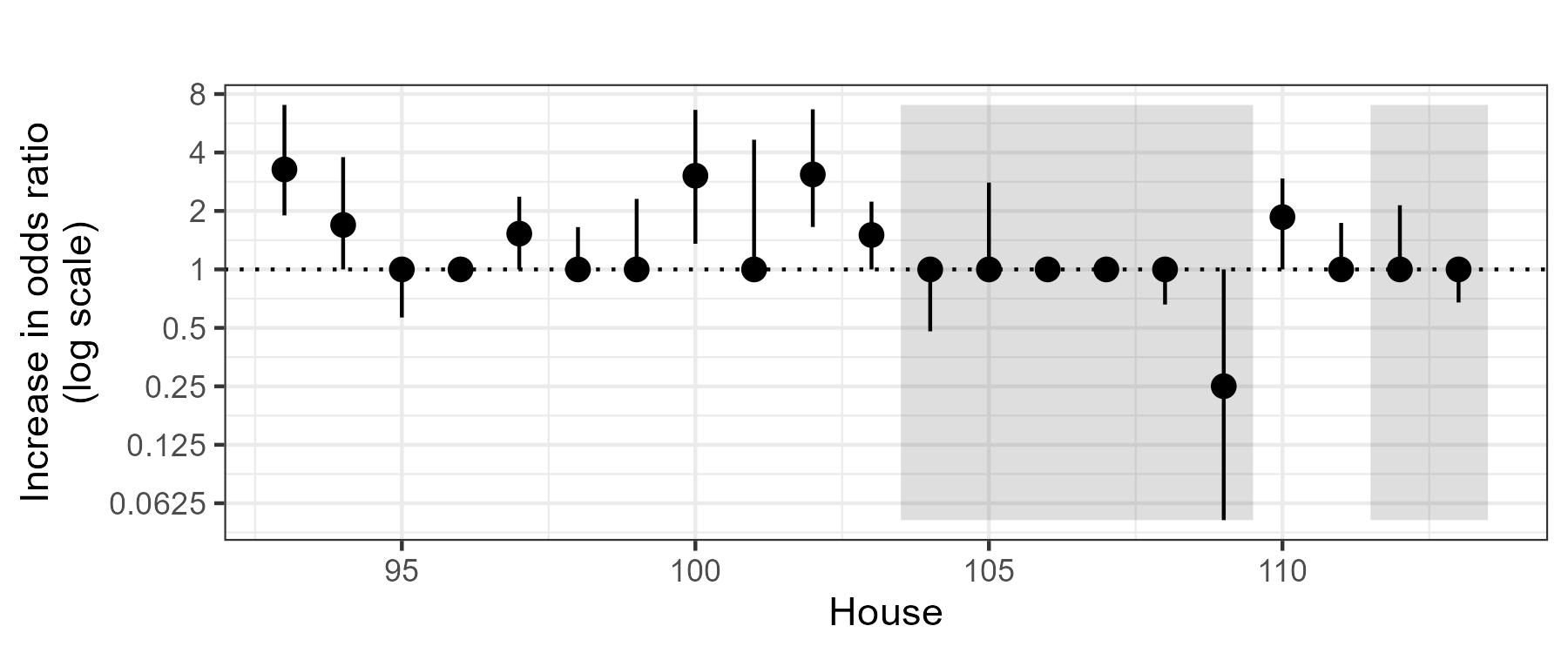}
  \caption{Influence of a $5\%$ increase in the proportion of constituency-level Republican vote in the most recent presidential election on odds of being a bridge legislator, for Republicans only.}\label{fig:prR}
  \end{subfigure}
\caption{Posterior median and 95\% percent credible intervals for the increase in odds ratios of being a bridge legislator for the three most important variables identified by our analysis.}
\label{fig:covariates}
\end{figure}

\begin{figure}[!ht]
\centering
\includegraphics[width=0.8\textwidth]{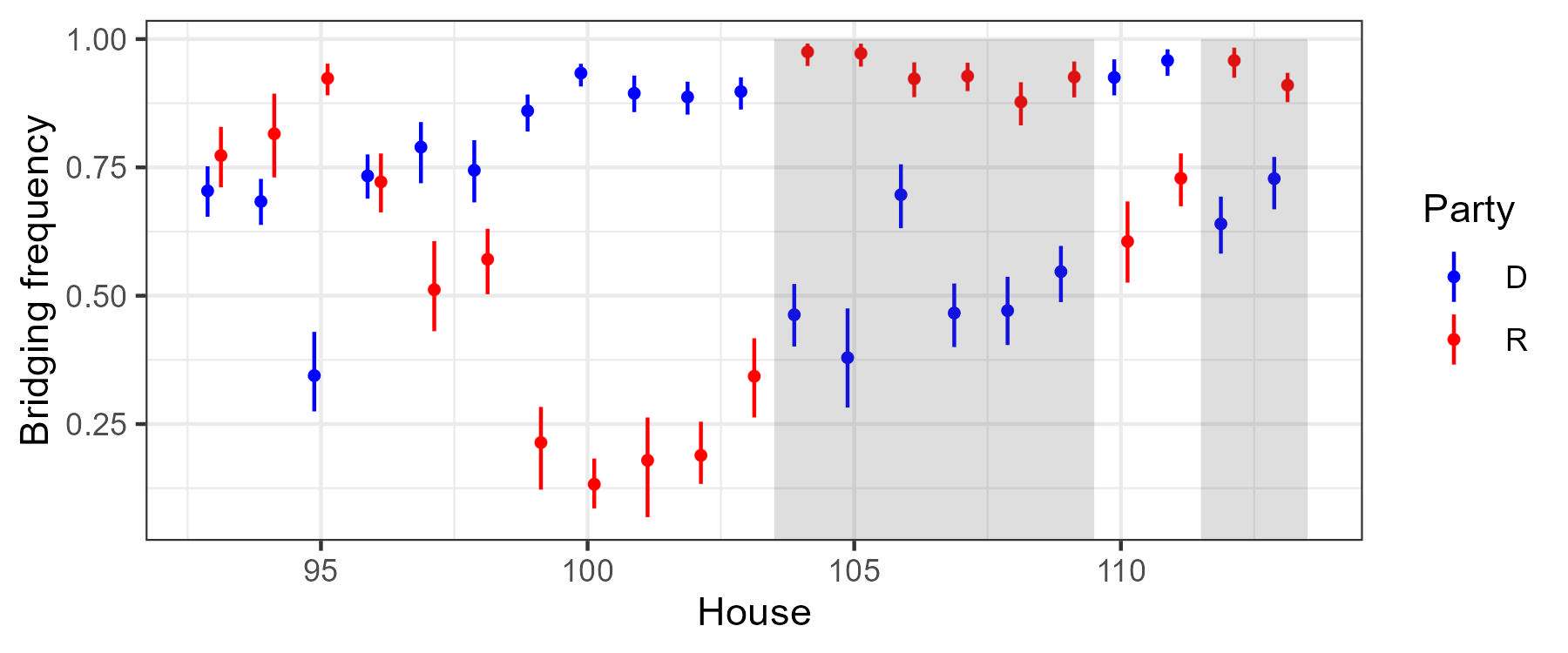}
\caption{Posterior mean and 95 percent credible intervals for the bridging frequency calculated separately for legislators in each party. Shaded regions represent sessions with a Republican majority and unshaded regions represent sessions with a Democratic majority.}
\label{fig:ASFparty}
\end{figure}

Figure \ref{fig:PIPs} shows the posterior inclusion probabilities (PIPs) for each covariate in each House under study.  The variable most commonly included in the model is \texttt{belto.partyControl}, the indicator of whether a given legislator belongs to the party in control of the House (its PIP is greater than 0.5 in 15 of the 21 Houses we studied, and is greater than 0.9 in all of these).  Except for the 95\textsuperscript{th} House, the coefficient associated with this variable is positive, with the model indicating that belonging to the party in control of the House increases the odds of being a bridge by a factor of anywhere between 3 and 435 (see Figure \ref{fig:beltoparty}). 
Another way to visualize this pattern is by reviewing raw bridging frequencies (recall Equation \eqref{eq:BF}) computed separately for each party (see Figure \ref{fig:ASFparty}).  Starting in 1985 (the 99\textsuperscript{th} House), there is a clear pattern in which the bridging frequency of the party in control of Congress is higher (and, in some cases, much higher) than the bridging frequency for the minority party.  This result should not be surprising in light of our previous discussions.  For the majority party to have any hope of passing legislation, it needs to achieve a certain level of party cohesion not only on procedural, but also on final passage votes.  Except perhaps in cases of very small majorities, the level of pressure on legislators from the minority party to toe the party line can be expected to be much lower \citep{pinn19-minority, rame15-bringing, robe05-minority}.

%
%

The next two variables that are most commonly included in the model correspond to \texttt{p\_R\_D} and \texttt{p\_R\_R}, the interactions between the party affiliation of a legislator and the percentage of their district's vote gathered by the Republican candidate in the most recent presidential election. The interaction associated with members of the Democratic party is included in the model in 9 of the 21 Houses under study and, in each of these cases, the associated coefficient is negative (indicating that, as their district leans more Republican, Democrats are less likely to be bridges).  If we exclude the 109\textsuperscript{th} congress (which is an outlier) and the 111\textsuperscript{th} congress (where this variable is only marginally included), a 5\% increase in the percentage of the Republican vote in the most recent presidential election decrease the odds of a Democratic legislator being a bridge by a factor of between 1.3 and 3.2 (see Figure \ref{fig:prD}).  On the other hand, the coefficient for members of the Republican party is included in the model for 8 Houses.  In 7 of those cases, the coefficient is now positive.  The exception is the 109\textsuperscript{th} House which, as we mentioned above, seems to be an overall outlier.  A 5\% increase in the percentage of the Republican vote in the most recent presidential election increases the odds of a Republican legislator being a bridge by a factor of between 1.5 and 3.3, roughly in line with what we observed for Democratic legislators (see Figure \ref{fig:prR}). 
Taken together, these two results suggests that legislators representing swing districts are more likely to vote differently in procedural and final passage votes.  This observation is consistent with the hypothesis that these legislators are subject to greater pressure from their constituents than legislators in safe districts.

None of the other variables included in our study seem to consistently explain bridging behavior by legislators.   If we leave out the outlier 109\textsuperscript{th} House, the number of cosponsored bills (\texttt{numCosp}) appears to be significant during the 95\textsuperscript{th} and the 108\textsuperscript{th} House, but the signs of the coefficients are opposite.  Similarly, the percentage of the population in the district that is unemployed (\texttt{prcntUnemp}) seems to be marginally significant in the 94\textsuperscript{th} and 107\textsuperscript{th} Houses, but again with different signs each time.  Finally the number of sponsored bills (\texttt{numSpon}), the indicator for whether a legislator is Hispanic (\texttt{hispanic}), the indicator for whether a legislator is Black (\texttt{black}), the percent of foreign born residents (\texttt{prcntForeignBorn}), and the percent of recent arrivals in the legislator's district (\texttt{recentArrivalPrcnt}) appear to be marginally significant in a single House each (the  111\textsuperscript{th},  94\textsuperscript{nd},  108\textsuperscript{nd},  95\textsuperscript{nd}, and  102\textsuperscript{nd},  respectively).

\section{Discussion}\label{se:discussion}

We have introduced a new class of hierarchical models that can be used to identify covariates that might explain why legislators vote differently across voting domains, with a focus on the differences in voting patterns between procedural and final passage votes in the U.S.\ House of Representatives.  The paper makes both methodological and applied contributions.  On the methodological side, our approach fully accounts for all the uncertainty associated with learning legislator's revealed preferences across the multiple domains.  The model is quite general and can be applied to a number of other settings (e.g., committee vs.\ floor voting, or voting on economic vs.\ social policy).  On the application side, our empirical results add to the literatures on competing principals in legislative voting and on substantive vs.\ procedural voting. Like \citet{cars14-procedural}, we find that party matters.  This is also consistent with \citet{flei04-shrinking,bond02-disappearance}.  We also find evidence that the constituency’s partisan leaning plays a role, which is consistent with \citet{jess14-two}.  In as much as this translates into electoral considerations, our results are also consistent with \citet{good04-lameduck}.  It is notable that we found that legislator-level characteristics such as gender and ethnicity do not seem to explain bridging behavior.  As far as we know, this is a novel insight, which is nonetheless in line with the results of \citet{schw04-gender}.

Our joint model seems to perform satisfactorily in most of the datasets we considered in Section \ref{se:example}.  The one possible exception is the 109\textsuperscript{th} House, where the model identifies a large number of variables as being potentially influential on the bridging probability.  In our discussion of the results, we attributed this outlier to the unique nature of the political scandals that arose during this period.  Indeed, while scandals of some sort are (sadly) not uncommon in the U.S.\ Congress, the nature of those ocurring during the 109\textsuperscript{th} House have a very particular significance in the context of our application, as they involved the highest echelons of the leadership of the party in control of the House.  Nonetheless, there are potential alternative explanations for this outlier.  One of them is model mis-specification.  Our hierarchical model can be conceived as being made of two ``modules'' (roughly speaking, one that learns the bridges, and one that determines which factors explain those bridges).  Over the last five years there has been growing interest on the impact that the mis-specification of one of the modules might have on overall model performance (e.g., see \citealp{jacob2017better} and \citealp{nott2023bayesian}).  In our case, the most likely misspecified module is the one that recovers the preferences of legislators from their voting records.  Indeed, recent evidence (e.g., see \citealp{yu2021spatial}, \citealp{duck2022ends} and \citealp{lei2023novel}) suggests that standard IRT models such as that in \textcite{jackman2001multidimensional} might fail to accurately capture the preferences of some legislators when those located at different extremes of the political spectrum tend to vote together against the more moderate legislators.  Solutions to potential mis-specifications issues include the use of more flexible models for capturing reveled preferences, as well the use of so-called ``cut'' inference \citep{plummer2015cuts,jacob2017better}.  These approaches will be explored elsewhere.

The approach developed in this paper is appropriate for situations in which voting happens across two domains.  A future area of research is how to extend the approach to situations in which there are more that two voting domains.  The most obvious approach (extending the latent logistic regression for binary data to a multinomial regression) quickly becomes impractical, even for a relatively small number of voting domains.  An alternative is to employ a covariate-dependent prior on partitions (e.g., see \citealp{muller2011product}, \citealp{dahl2017random} or \citealp{page2018calibrating}) to implicitly define a prior distribution on the probability that legislators reveal the same preferences across any pair of voting domains.  This kind of extension will be explored elsewhere.

\begin{acknowledgement}
We would like to thank Stephen Jessee and Sean Theriault for providing access to their data.
\end{acknowledgement}

\paragraph{Funding Statement}

This research was supported by grants from the National Science Foundation,  2114727 and 	2023495.

\paragraph{Competing Interests}

None.


\appendix

\section{Explanatory variables considered}\label{ap:variable}

Demographic and socioeconomic data for the legislators and constituencies was obtained from \textcite{foster2017historical}.  Data on the results of the presidential elections between 1970 and 2008 was generously provided by Stephen Jessee and Sean Theriault (personal communication), while data for the 2012 and 2016 elections was obtained from 
\textcite{KosReport}.

\begin{enumerate}

\item Covariates related to party affiliation

\begin{description}
\item[belto.partyControl:] Indicator for whether the legislator is a member of the party that has the majority in the current House.
\item[p\_R\_R and p\_R\_D:] These two covariates are interactions between the two party membership indicators and a measure of partisan political ideology for the legislator's district. The Republican share of the two-party presidential vote (p\_R) is the percentage of the two-party vote won by the Republican candidate in the most recent presidential election (centered so that $0$ indicates that the two parties received an equal percentage of the vote). $p\_R\_R=I(Republican)\times p\_R$ is equal to the Republican voteshare $p\_R$ for legislators belonging to the Republican party and $0$ for legislators belonging to the Democratic party. The other interaction $p\_R\_D=I(Democrat)\times p\_R$ is defined an analogous way. 
\end{description}

\item Legislator characteristics
\begin{description}
\item[age:] Age at time of being sworn into congress for current session.
\item[gender:] Gender of legislator.
\item[black:] Indicator for membership to the Congressional Black Caucus. The authors of the data note that to their knowledge, all self-identifying Black members of congress are members of the caucus.
\item[hispanic:] Indicator for membership to the Congressional Hispanic Caucus. The authors of the data note that to their knowledge, all self-identifying Hispanic members of congress are members of the caucus.
\item[numberTerms:] Number of terms served in the House.
\item[numSpon:] Number of bills sponsored by the legislator in the current term. 
\item[numCosp:] Number of bills co-sponsored by the legislator in the current term.
\item[numPassH:] Number of bills sponsored by the legislator that were approved by a full House vote in the current term.
\end{description}

\item Constituency characteristics (based on data from Census)

\begin{description}
\item[recentArrivalPrcnt:] Percentage of the district that recently moved to the district from another county (note that the census does not track how many people have moved into a district from within the same county).
\item[prcntForeignBorn:] Percentage of the district that was born in a foreign country.
\item[gini:] Index of economic inequality calculated based on the percentage of the population in each income bracket.
\item[ses:] Measure of socioeconomic status calculated based on the income and education level of the district. 
\item[prcntUnemp:] Percentage of the district's population that is unemployed but still in the labor force.
\item[prcntBA:] Percentage of the district with a bachelor's degree or higher.
\item[prcntHS:] Percentage of the district with a high school degree or higher.
\item[prcntBlack:] Percentage of the district that is Black, including those who are Black and Hispanic.
\item[prcntHisp:] Percentage of the district that is Hispanic (both Black and White).
\item[prcntAsian:] Percentage of the district that is Asian.
\end{description}

\end{enumerate}

\printbibliography

\end{document}


\maketitle

\section{Details of MCMC sampler}

\begin{enumerate}

\item Sample parameters associated with the policy space

\begin{enumerate}
\item Sample missing observations. For $i=1,\dots,I$ and $j=1,\dots,J$, if $y_{i,j}$ is missing, sample 
\[y_{i,j} \mid \cdots \sim \Ber \left( \, \frac{\exp\left\{ \mu_j+\bfalp_j^T\bfbet_{i,\gam_j}\right\}}{1 + \exp \left\{ \mu_j+\bfalp_j^T\bfbet_{i,\gam_j} \right\}} \right).\]

\item Sample latent Polya-Gamma variables.
For $i=1,\dots,I$ and $j=1,\dots,J$, sample
\[\nu_{i,j}\mid \cdots \sim PG(1,\mu_j+\alf_j\bet_{i,\gam_j}).\]
In what follows, let 
$z_{i,j}\equiv \frac{y_{i,j}-1/2}{\nu_{i,j}}$.

\item Sample parameters associated with policy space

\begin{enumerate}
\item Sample $\mu_j$. For $j=1,\dots,J$, sample $\mu_j\mid \cdots\sim N(\hat\rho_\mu, \hat\kappa^2_\mu)$, where
\[\hat\kappa^2_\mu
=\left[\frac{1}{\kappa_\mu^2} + \sum_{i=1}^I \nu_{i,j}\right]^{-1}
\text{ and }
\hat\rho_\mu = \hat\kappa^2_\mu
\left[\frac{\rho_\mu}{\kappa^2_\mu}
+\sum_{i=1}^I \nu_{i,j}(z_{i,j}-\alf_j\bet_{i,\gamma_j})\right].\]

\item sample $\alf_j$
For $i=1,\dots,I$, sample 
$\alf_j\mid\cdots \sim \hat\omega_\alf\del_0(\alf_j)+(1-\omega_\alf)N(\hat\rho_\alf,\hat\kap^2_\alf)$, where

\[\hat\kappa^2_\alf
=\left[\frac{1}{\kappa_\alf^2} + \sum_{i=1}^I \nu_{i,j}\bet_{i,\gam_j}^2\right]^{-1}
\text{ and }
\hat\rho_\alf = \hat\kappa^2_\alf
\left[\sum_{i=1}^I \nu_{i,j}(z_{i,j}-\mu_j)\bet_{i,\gam_j}\right]\]
and
\[\frac{\hat\omega_\alf}{1-\hat\omega_\alf} = \frac{\omega_\alf}{1-\omega_\alf}\cdot \frac{\kap_\alf}{\hat\kap_\alf}\cdot 
\exp\left\{-\frac{\hat\rho^2_{\alf}}{2\hat\kap^2_{\alf}}\right\}.\]

\item Sample $\zeta_i$. For $i=1,\dots,I$, sample $\zeta_i$ according to $P(\zeta_i=1\mid\cdots)=\frac{p_1}{1+p_1}$, where

\[p_1
=\frac{\tht_i}{1-\tht_i}
\sqrt\frac{\sig^2_\bet\hat\sig^2_\bet}{\hat\sig^2_{\bet_0}\hat\sig^2_{\bet_1}}
\frac{\exp\left(\hat\sig^2_\bet
\left[\sum_{j=1}^J q_{i,j}\alf_j\nu_{i,j}\right]^2\right)}
{\exp\left(\hat\sig^2_{\bet_0}
\left[\sum_{j:\gam_j=0} q_{i,j}\alf_j\nu_{i,j}\right]^2\right)
\exp\left(\hat\sig^2_{\bet_1}
\left[\sum_{j:\gam_j=1}q_{i,j}\alf_j\nu_{i,j}\right]^2\right)},\]

where $q_{i,j}=z_{i,j}-\mu_j-\eta_\bet \alf_j,$
\[\hat\sig^2_\bet=\left(\frac{1}{\sig^2_\bet}+
\sum_{j=1}^J \nu_{i,j}\alf_j^2\right)^{-1}
\text{ and }
\hat\sig^2_{\bet_\gam}=\left(\frac{1}{\sig^2_\bet}+
\sum_{j:\gam_j=\gam} \nu_{i,j}\alf_j^2\right)^{-1}.\]
Here $\tht_i=\frac{e^{\eta_0+\bfx_i^T\bfeta}}{1+e^{\eta_0+\bfx_i^T\bfeta}}$ is the probability from the covariate regression.

\item Sample $\beta_{i,\gam}$. For $i=1,\dots,I$, sample
\[
\beta_{i,0},\beta_{i,1}|\cdots \sim
\begin{cases}
N(\beta_{i,0}\mid \hat\rho_\bet,\hat\sig^2_\bet)
\del_{\bet_0}(\bet_{i,\gam})
& \text{ if }\zeta_i=1 \\
N(\beta_{i,0}\mid \hat\rho_{\bet_0},\hat\sig^2_{\bet_0})
N(\beta_{i,1}\mid \hat\rho_{\bet_1},\hat\sig^2_{\bet_1})
& \text{ if }\zeta_i=0 \\
\end{cases}
,\]

where 
\[\hat\rho_\bet
=\hat\sig^2_\bet
\left[\frac{\rho_\bet}{\sig^2_\bet}
+\sum_{j=1}^J \nu_{i,j}(z_{i,j}-\mu_j)\alf_j\right]
\text{ and }
\hat\rho_{\bet_\gam}
=\hat\sig^2_{\bet_\gam}
\left[\frac{\rho_\bet}{\sig^2_\bet}
+\sum_{j:\gam_j=\gam} \nu_{i,j}(z_{i,j}-\mu_j)\alf_j\right].\]

\end{enumerate}

\item Enforce identifiability
\item Sample hyperparameters
\end{enumerate}

\item Sample parameters associated with explaining identity of bridges

\begin{enumerate}
\item Sample latent Polya-Gamma variables
For $i=1,\dots,I$, sample
\[\nu_i\mid \cdots \sim PG(1,\eta_0+\bfx_i^T\bfeta).\]
In what follows, let 
$z_i\equiv \frac{\zeta_i-1/2}{\nu_i}$.

\item Sample intercept

\item Sample model. To sample the model $\bfxi$, we use a Metropolis-Hastings sampler with the following proposal:
\begin{itemize}
    \item With probability $p_1=0.9$, flip $k$ elements of $\bfxi$ from $0$ to $1$ or vice versa. Choose $k\in\{1,2,3,4\}$ with probabilities $p_2=0.6, 0.2, 0.15, 0.05$. Given $k$, choose the set of elements to flip uniformly at random.
    \item With probability $1-p_1=0.1$, remove one covariate from the model and add another covariate, keeping the model size the same. Choose the covariates to add and remove uniformly at random. If $\sum_{q=1}^p \bfxi_q\in\{0,p\}$ (the null or saturated model), this second proposal type is not possible so we always select the first type of proposal.
\end{itemize}

The above proposal is reversible, and the acceptance probability is given by
\[A(\xi,\xi^*)
=min\left\{1,\frac{BF(\bfxi^*,\bfxi_0|\bfzeta,\eta_0)}
{BF(\bfxi,\bfxi_0|\bfzeta,\eta_0)}
\times \frac{{p\choose p_{\bfxi}}}{{p\choose p_{\bfxi^*}}},
\right\}\]
where $p_{\bfxi}=\sum_{q=1}^p \bfxi_q$ and $BF(\bfxi,\bfxi_0|\bfzeta,\eta_0)$ is the Bayes factor for comparing $\bfxi$ to the null model $\bfxi_0=\bfzero_p$.

The Bayes factor is given by
\begin{align*}
BF(\bfxi,\bfxi_0|\bfzeta,\eta_0)
&=\sqrt\frac{1}{|\bfI+4I\bfX_{\bfxi}^T\bfV\bfX_{\bfxi}(\bfX_{\bfxi}^T\bfX_{\bfxi})^{-1}|}
\\ &\qquad\qquad
\exp\left\{\frac{1}{2}(\bfz-\eta_0)^T
\bfV \bfX_{\bfxi}
\left[\bfX_{\bfxi}^T\left(
\frac{1}{4I}\mathbf{I} + \bfV 
\right)\bfX_{\bfxi} \right]^{-1}
\bfX^T_{\bfxi} \bfV(\bfz-\eta_0)
\right\},
\end{align*}
where $\bfV=diag(\nu_1,\dots,\nu_I)$ and $\bfI$ is the identity matrix with dimensions $p_{\bfxi}$,

\item Sample coefficients, Sample $\bfeta\mid\cdots 
\sim N(\hat\mu, \hat\Sig)$,
where 
\[\hat\Sig = \bfX_{\bfxi}^T\left(V+\frac{1}{4I}I_{p^*}\right)\bfX_{\bf\xi}
\text{ and }
\hat\mu = \hat\Sig\cdot \bfX_{\bfxi}^T\bfV(\bfz-\eta_0)^T.\]

\end{enumerate}

\end{enumerate}